\documentclass[runningheads,pdftex]{llncs}

\usepackage[crop,final,accepted]{espositollncs}
\conference{IS-EUD 2025: 10th International Symposium on End-User Development}

\usepackage[T1]{fontenc}
\usepackage{graphicx}
\usepackage{hyperref}
\usepackage{color}

\usepackage[nolist]{acronym}
\usepackage[capitalise]{cleveref}
\usepackage{booktabs}
\usepackage{tabularx}
\usepackage{float}
\usepackage{subcaption}
\usepackage[inline]{enumitem}

\usepackage{soul}

\begin{document}
\title{Explanation-Driven Interventions for Artificial Intelligence Model Customization}
\subtitle{Empowering End-Users to Tailor Black-Box AI in Rhinocytology}
\titlerunning{Explanation-Driven Interventions for AI Model Customization}
%
\author{Andrea Esposito\inst{1}\corresponding\orcidID{0000-0002-9536-3087} \and
Miriana Calvano\inst{1}\orcidID{0000-0002-9507-9940} \and
Antonio Curci\inst{1,2}\orcidID{0000-0001-6863-872X} \and
Francesco Greco\inst{1}\orcidID{0000-0003-2730-7697} \and
Rosa Lanzilotti\inst{1}\orcidID{0000-0002-2039-8162} \and
Antonio Piccinno\inst{1}\orcidID{0000-0003-1561-7073}%
}
\authorrunning{A. Esposito et al.}
%
\institute{%
Department of Computer Science, University of Bari Aldo Moro, Via E. Orabona 4, 70125, Bari, Italy\\
\email{\{andrea.esposito,miriana.calvano,antonio.curci,\\francesco.greco,rosa.lanzilotti,antonio.piccinno\}@uniba.it}%
\and%
Department of Computer Science, University of Pisa, Largo B. Ponte Corvo, 3, Pisa, 56127, Italy
}
\maketitle              
\begin{abstract}
The integration of Artificial Intelligence (AI) in modern society is transforming how individuals perform tasks. In high-risk domains, ensuring human control over AI systems remains a key design challenge. This article presents a novel End-User Development (EUD) approach for black-box AI models, enabling users to edit explanations and influence future predictions through targeted interventions. By combining explainability, user control, and model adaptability, the proposed method advances Human-Centered AI (HCAI), promoting a symbiotic relationship between humans and adaptive, user-tailored AI systems.

\keywords{Customization  \and Black-box AI \and Model Reconfiguration \and Explainable AI (XAI) \and Human-AI symbiosis}
\end{abstract}
\acresetall

\begin{acronym}
	\acro{EUD}{End-User Development}
	\acro{AI}{Artificial Intelligence}
	\acro{HCAI}{Human-Centered Artificial Intelligence}
	\acro{XAI}{eXplainable Artificial Intelligence}
    \acro{SAI}{Symbiotic Artificial Intelligence}
	\acro{UI}{User Interface}
\end{acronym}
\section{Introduction}\label{sec:introduction}

\ac{AI} has become an integral component of decision-support systems in numerous domains, including healthcare, finance, and law \cite{Guidotti2019Survey}. While \ac{AI} models could enhance decision-making, their reliance on complex, often opaque, algorithms presents a significant barrier to adoption \cite{Guidotti2019Survey}, especially in high-stakes applications such as medical diagnostics \cite{Combi2022Manifesto}. End-users---typically domain experts rather than \ac{AI} specialists---require mechanisms to refine and adjust \ac{AI} behavior to better align with their expertise and contextual knowledge.

Most AI systems follow a one-size-fits-all approach, with limited support for post-deployment customization \cite{Chambers2022Creating}. This lack of adaptability can lead to misaligned recommendations, loss of trust, and decreased usability. \ac{EUD} for \ac{AI} seeks to address this issue by enabling non-technical users to customize \ac{AI} behavior according to their specific needs \cite{Fischer2023Adaptive}.

While the integration of \ac{EUD} for \ac{AI} has made substantial progress in fields such as the Internet of Things (IoT), education, and business analytics \cite{Esposito2023EndUser,Li2022How}, \ac{AI}-based decision-support systems---especially those powered by black-box models---remain challenging for users to modify. Current approaches to \ac{AI} customization for end-users are
\begin{enumerate*}[label=(\roman*)]
\item \emph{Rule-Based Customization} \cite{Esposito2023EndUser},
\item \emph{Low-Code / No-Code \ac{AI}} \cite{Li2022How}, and
\item \emph{Human-\ac{AI} Collaboration Interfaces} \cite{Desolda2024Humancentered,Raees2024Explainable}.
\end{enumerate*}
The first one involves the definition from the user of if-this-then-that conditions to influence \ac{AI} outputs. 
On the other hand, the platforms that adopt \emph{Low-Code / No-Code \ac{AI}} allow users to build and deploy \ac{AI} models without programming (e.g., AutoML tools), but they do not enable real-time intervention on model behavior post-deployment. Lastly, \emph{Human-\ac{AI} Collaboration Interfaces} enable users to validate and/or override the system's predictions without affecting its future behavior. These approaches do not fully address the need for an interactive and iterative refinement of \ac{AI} behavior based on human expertise. Using professionals' knowledge and expertise to refine \ac{AI} reasoning can be a valuable resource to improve the system's performance while building a stronger symbiotic relationship between the two parties \cite{Grigsby2018Artificial,Desolda2024Humancentered}.

Creating \ac{AI}-based systems that embody these characteristics can foster collaboration, establishing a symbiotic relationship between humans and \ac{AI}. \ac{SAI} is a specialization of Human-Centered \ac{AI} \cite{Shneiderman2022HumanCentered} and aims at supporting humans instead of replacing them. This implies creating solutions that reflect humans' needs and preferences by integrating intervention paradigms, transparency, and fairness by design focusing on augmentation rather than automation \cite{Jarrahi2018Artificial,Desolda2024Humancentered}.

This research proposes a novel intervention-based \ac{UI} within the \emph{Rhino-Cyt} platform, designed to empower rhinocytologists to modify \ac{AI}-generated classifications and explanations. This proposal aims at embodying the \emph{EUDability} construct \cite{Barricelli2023EUDability} by introducing an innovative explanation-driven \ac{EUD} approach, allowing end-users to adjust \ac{AI} classifications, edit \ac{AI}-generated explanations, and indirectly refine and tailor the \ac{AI} model through the mechanism of \textit{interventions} \cite{Schmidt2017Intervention}. This approach goes beyond rule-based or component-based customization, offering a \emph{human-in-the-loop} model refinement paradigm. Thus, Rhino-Cyt involves rhinocytologists as its end-user developers, supporting them in reaching the goal of diagnosing.

The rest of the article is structured as follows. \cref{sec:background} discusses prior research in \ac{EUD} for AI, explainability, and human-\ac{AI} collaboration, also presenting Rhino-Cyt. \cref{sec:ui} presents the design of the intervention-based \ac{UI}, detailing its interaction flow and impact on \ac{AI} adaptation. \cref{sec:comparison} positions Rhino-Cyt within existing \ac{EUD} for \ac{AI} taxonomies and compares it with other customization paradigms. \cref{sec:conclusions} concludes this article by summarizing its key contributions and outlining the next steps.

\section{Background and Related Work}\label{sec:background}

This section reviews prior research on \ac{EUD} for \ac{AI}, explainability as a mechanism for \ac{EUD}, and human-\ac{AI} collaboration in decision-support systems.

\subsection{End-User Development for AI}

\ac{EUD} encompasses a range of methods, techniques, tools, and socio-technical environments that empower non-professionals to engage in activities usually reserved for professionals in ICT-related areas, including the ability to create, modify, extend, and test digital artifacts without requiring specialized knowledge in conventional software engineering practices \cite{Barricelli2019Enduser}. A systematic literature review by Esposito et al. categorized existing \ac{EUD} for \ac{AI} approaches into five key paradigms \cite{Esposito2023EndUser}:
\begin{enumerate}
	\item \emph{Component-Based:} Users assemble predefined \ac{AI} components through visual programming interfaces.
	\item \emph{Rule-Based:} Users are allowed to modify \ac{AI} behavior through ``if-then'' rules.
	\item \emph{Wizard-Based:} Step-by-step guidance simplifies \ac{AI} customization, presenting the task as a sequence of operations that guide users throughout the overall activity.
	\item \emph{Template-Based:} Users adjust pre-built \ac{AI} models by modifying parameters.
	\item \emph{Workflow and Data Diagrams: }Users define \ac{AI} processes using flow-based representations.
\end{enumerate}


While these approaches are effective for tasks such as building \ac{AI} models from scratch or configuring predefined automation rules, they offer limited support for modifying existing black-box \ac{AI} models.
\emph{Explanation-driven interventions} extend \ac{EUD} for \ac{AI} by introducing a new paradigm. Instead of requiring users to manipulate \ac{AI} model components or logic directly, this proposal allows them to edit \ac{AI}-generated explanations, \emph{indirectly} refining the model's behavior over time.


In this regard, explainability has rarely been explored as an active mechanism for \ac{EUD}. 
\Ac{XAI} seeks to make \ac{AI} model decisions more interpretable and transparent \cite{Guidotti2019Survey,Sahoh2023Role}. Through \ac{XAI}, users can understand \ac{AI} decisions, deciding whether to rely on its predictions or potentially recognize (and mitigate) biases by assessing its reasoning \cite{Guidotti2019Survey}.

In most \ac{AI}-assisted decision systems, explanations are static, not allowing users to modify them to influence future \ac{AI} behavior. Our proposal introduces a novel editable explanation mechanism where users can modify explanations associated with \ac{AI} classifications, provide domain-specific refinements to ensure \ac{AI}-generated explanations align with expert knowledge, and influence future \ac{AI} behavior indirectly.

\subsection{Human-\ac{AI} Collaboration in Decision Support Systems}

\ac{AI}-based decision-support systems, especially in medicine, often follow a \emph{human-on-the-loop} paradigm, where users oversee the decision-making process by interacting with \ac{AI} outputs to merely validate its decisions \cite{Sherson2023MultiDimensional,Desolda2024Humancentered,Fischer2021Loop}. This translates into a validation-based collaboration, where experts review \ac{AI} predictions but have no direct manipulation mechanism for modifying the \ac{AI}'s reasoning process \cite{Desolda2024Humancentered}. 

Our proposal aims at filling this gap following the model of human--\ac{AI} interaction proposed by Desolda et al. \cite{Desolda2024Human}, moving beyond this by enabling direct, explanation-driven interventions, ensuring that \ac{AI}'s decision-making processes evolve alongside domain experts, fostering a symbiosis among humans and AI \cite{Desolda2024Humancentered}. Establishing proper collaboration mechanisms between humans and \ac{AI} is crucial to guarantee that professionals are aware of the processes that lead to outputs. 


\subsection{Rhino-Cyt: an AI-based system for Rhinocitology}

Our proposal leverages a case study in the context of rhinocytology (a subfield of medical cytology) \cite{Dimauro2019Nasal,Dimauro2018RhinoCyt}. Currently, the diagnostic process in rhinocytology is mainly based on direct observation under the microscope, which requires a prolonged effort by rhinocytologists \cite{Gelardi2012Atlas,Desolda2024Human}. 

Rhino-Cyt is an \ac{AI}-assisted environment for the classification of nasal cytology samples that supports its end-user developers, i.e., medical professionals, in diagnosing inflammatory and allergic conditions \cite{Dimauro2018RhinoCyt}. It employs AI models (namely, a CNN) to automate the cytological examination by segmenting histological samples of the nasal mucosa, identifying and classifying individual cells based on nine cytotypes \cite{Dimauro2018RhinoCyt,Gelardi2012Atlas}: (i) ciliated, (ii) muciparous, (iii) basal cells, (iv) striated cells, (v) neutrophils, (vi) eosinophils, (vii) mast cells, (viii) lymphocytes, (ix) metaplastic cells.

\section{Design of the Intervention-Based User Interface}\label{sec:ui}
\begin{figure}[h!]
		\centering
		\includegraphics[width=0.7\linewidth]{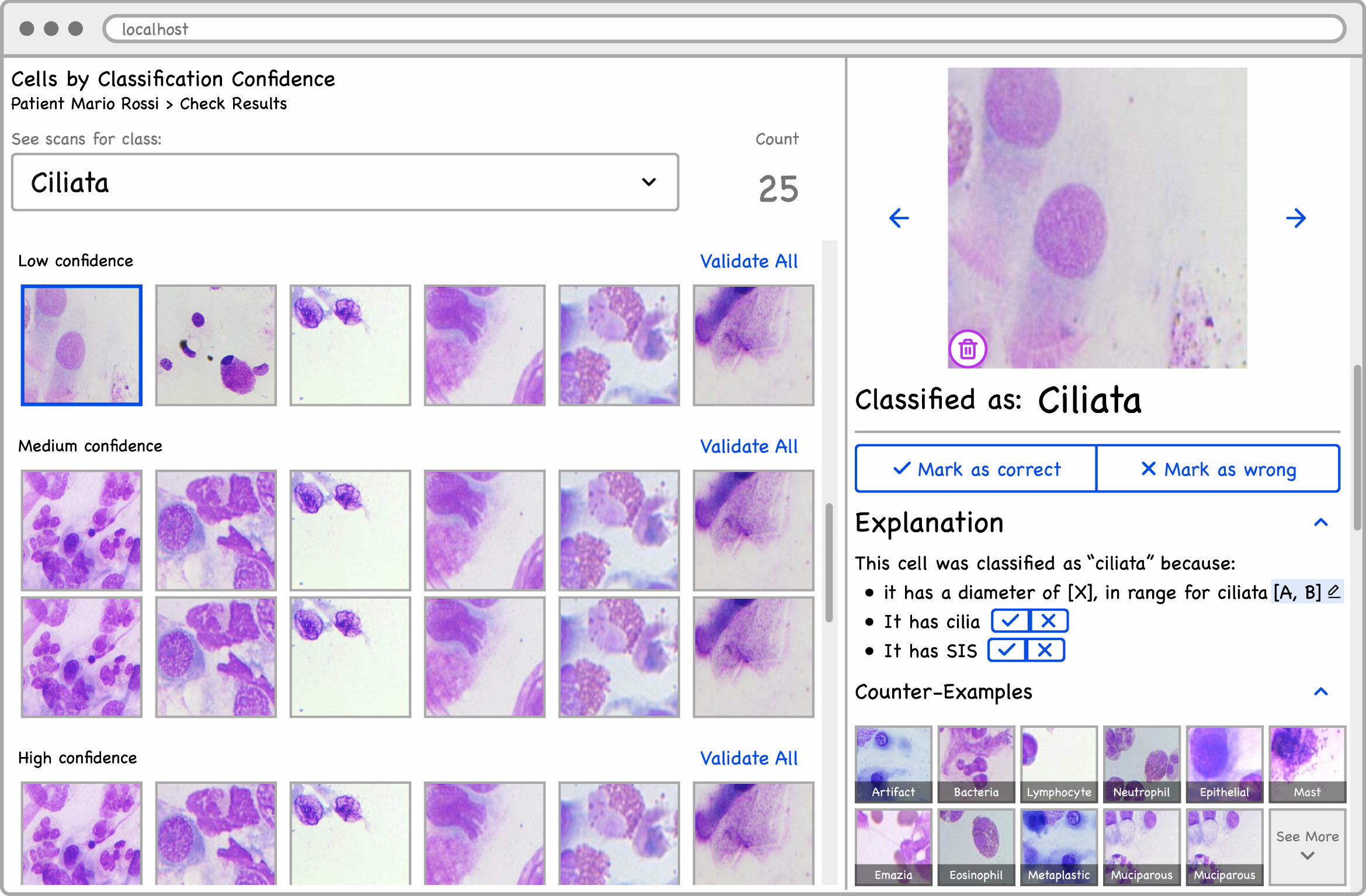}
        \label{fig:rhinocyt-ui:ui}
	\caption{The user interface for allowing black-box \ac{AI} model tailoring implemented in Rhino-Cyt}
	\label{fig:rhinocyt-ui}
\end{figure}

Rhino-Cyt introduces an innovative intervention-based \ac{UI} that allows rhinocytologists to refine \ac{AI}-assisted classifications, presented in \cref{fig:rhinocyt-ui}. Users can \emph{intervene} modifying decisions and explanations, unlike conventional \ac{AI} customization methods, which rely on rule-based or low-code paradigms \cite{Schmidt2017Intervention}.

This section details the design principles, \ac{UI} components, and interaction workflow, demonstrating how the system aligns with \ac{EUD} for \ac{AI} by offering an explanation-driven intervention mechanism.

\subsection{Design Rationale and Principles}

The design of the Rhino-Cyt intervention \ac{UI} follows three core principles derived from \ac{EUD} for \ac{AI} research \cite{Esposito2023EndUser}, which emphasize the human intervention that leads to the \ac{AI} customization while minimizing humans' cognitive load. The referred principles are:
\begin{enumerate}
    \item \emph{Intervention-Based Interaction}. The system allows logged users to adjust \ac{AI} classifications and explanations tracking any modifications. 
    \item \emph{Explainability-Driven Customization}. The system supports it by allowing users to refine \ac{AI}-generated justifications without requiring them to modify raw model parameters or write explicit rules. Instead, the interface leverages editable explanations as a means of customization, enabling domain experts to adjust and improve \ac{AI}-generated reasoning based on their medical expertise. 
    \item \emph{Minimal Cognitive Load for Domain Experts}. The system is designed to minimize the cognitive load for domain experts by providing guided interventions that simplify the interaction process and reduce the need for technical expertise. \ac{AI} adaptation occurs implicitly through user feedback, allowing the model to refine its reasoning without requiring manual retraining.
\end{enumerate}


	

\subsection{User Interface Components}

The elements of Rhino-Cyt's \ac{UI} (\cref{fig:rhinocyt-ui}) that allow users to intervene on \ac{AI} decisions consist of the following key components
\begin{enumerate}
    \item The \emph{Classification Details and Interventions Panel} presents AI-generated classifications, allowing users to review system decisions. If necessary, users can override \ac{AI}'s decisions by selecting an alternative category which is incorporated into the system, contributing to the continuous adaptation and improvement of the \ac{AI}'s future performance.
    
    \item The \emph{Editable Explanation Area} displays the \ac{AI}-generated justifications of its classifications, providing transparency into the decision-making process. Users can modify these justifications to guarantee they align more closely with their expertise. Any edits made by users directly influence the AI's reasoning model, refining its approach and shaping future justifications.
\end{enumerate}

\subsection{Interaction Workflow: How Users Intervene in \ac{AI} Decisions}

The Rhino-Cyt intervention workflow is structured as a three-step process.

\begin{enumerate}[label={\itshape Step \arabic*:},align=left,labelsep=5pt,widest={Step 10:}]
    \item \emph{Reviewing \ac{AI}-Generated Classification and Explanation.} Upon analyzing a nasal cytology sample, the \ac{AI} presents a predicted classification and a textual explanation justifying the classification.
    
    \item \emph{User Intervention via Adjustment or Explanation Editing.} The users have two options for intervention: they can either override the \ac{AI} classification by selecting an alternative label or modify the \ac{AI}-generated explanation to reflect expert reasoning more accurately, or both.
    
    \item \emph{Model Adaptation and Visualization of Impact.} The system logs interventions and updates the \ac{AI}'s explanation model.
\end{enumerate}

This workflow enables \ac{AI} adaptation to be progressive, ensuring the model evolves alongside experts' knowledge, thus creating a human-in-the-loop customization mechanism for \ac{AI}-supported decision-making.

\subsection{Underlying \ac{AI} Model Adaptation}

Our proposal balances manual user intervention with an automated model refinement process based on users' feedback, thus providing an example of both an \emph{adaptable} and \emph{adaptive} system \cite{Fischer2023Adaptive}. 

The platform continuously logs user interactions, capturing every action performed by users. This includes explicit feedback on the \ac{AI} model's predictions, where users can either accept the suggested classification or reject it and make modifications. The collected feedback is then leveraged to initiate the retraining of the \ac{AI} model.

To enhance the adaptivity and adaptability of the \ac{AI} system, several techniques can be employed, ranging from online learning \cite{Amari1967Theory,Tallec2018Unbiased} to reinforcement learning from human feedback \cite{Kaufmann2023Survey}, leveraging detailed explanation obtained either from a gray-box model (as the one presented by Desolda et al.~\cite{Desolda2024Human}) or through large-language models. Specifically, user feedback on explanations helps refine the \ac{AI} model in different ways. If there is a direct, one-to-one relationship between an explanation and the model's decision-making process---as seen in decision trees---the model can be updated immediately. In contrast, when using more complex models, such as deep learning, the feedback is transformed into additional data points that contribute to adjusting the model's internal parameters.

\section{Positioning Explanation-Driven Interventions as an EUD Tool for AI}\label{sec:comparison}

The Rhino-Cyt intervention \ac{UI} introduces a novel approach to \ac{EUD} for \ac{AI}, using editable explanations as an \ac{AI} customization mechanism. This section explores Rhino-Cyt's classification within the current \ac{EUD} for \ac{AI} landscape, comparing it with existing AI customization approaches while also examining its impact on human-AI symbiosis. Rhino-Cyt's main features are summarized in \cref{tab:summary} and described below to highlight the customization aspects of the interaction workflow. 

\begin{table}[h!]
	\centering
	\caption{A summary of the main features of Rhino-Cyt's intervention-based \ac{UI}}
	\label{tab:summary}
	\begin{tabular}{@{}ll@{}}
		\toprule
		Feature & Rhino-Cyt's Intervention \ac{UI} \\ \midrule
		Customization Approach & Explanation-driven intervention \\
		User Control & Freeform modifications of \ac{AI} outputs and justifications\\
		Impact on Model Behavior & Direct, real-time adaptation \\
		Technical Expertise Required & None (domain expertise only) \\
		\bottomrule
	\end{tabular}
\end{table}

It enables end-users to customize its functionalities through explanation-driven intervention.
Based on the information highlighted in the explanation, users can take actions such as marking the explanation as accurate or incorrect or adjusting the feature values used in the reasoning process. This approach puts physicians in control by allowing them to refine the system's outputs and justifications, enhancing its performance, especially in cases involving outliers. In this context, user actions can represent feedback for the system that can adapt its behavior over time.

To classify Rhino-Cyt within the existing \ac{EUD} for \ac{AI} landscape, we adopt a recent framework proposed by Esposito et al. \cite{Esposito2023EndUser}, which categorizes \ac{EUD} \ac{AI} solutions based on the dimensions presented in \cref{tab:classification}.

\begin{table}[t]
	\centering
	\caption{Classification of the intervention-based \ac{UI} according to \cite{Esposito2023EndUser}}
	\label{tab:classification}

\begin{tabular}{@{}ll@{}}
\toprule
EUD Dimension & Rhino-Cyt Implementation \\ \midrule
Composition Paradigm & Explanation-driven, rule-based intervention \\
Target Users & Domain experts (rhinocytologists) \\
Technology & AI-assisted medical diagnostics \\
Usage & Single-user, with potential for collaborative interventions \\
Customization Level & Tailoring and indirect model refinement \\
Approach Output & AI model adaptation via explanation modifications \\
\bottomrule
\end{tabular}
\end{table}



To illustrate the novelty of Rhino-Cyt, we compare it with three common \ac{AI} customization paradigms: rule-based customization, no-code model building, and collaboration interfaces. The comparison is reported in \cref{tab:comparison}.

\begin{table}[t]
	\centering
	\caption{Comparison between the three common customization paradigms and the Rhino-Cyt intervention-based \ac{UI}.}
	\label{tab:comparison}

\begin{tabularx}{\linewidth}{@{}XXp{1.75cm}X@{}}
	\toprule
	Approach & Customization Scope & Technical Expertise Required & Impact on \ac{AI} Model \\ \midrule
Rule-Based \ac{AI} Customization & Predefined rule sets & Moderate & Direct, deterministic changes \\
No-Code \ac{AI} Model Building & Component-based visual programming & Low & Configures \ac{AI} before deployment \\
Human-\ac{AI} Collaboration Interfaces & Users validate/override \ac{AI} outputs & None & No direct \ac{AI} adaptation \\ \midrule
Explanation-Driven Intervention & Editable explanations influence \ac{AI} reasoning & None & Indirect, adaptive refinements over time \\
	\bottomrule
\end{tabularx}
\end{table}

\textit{Explanation-Driven Intervention} enables users to manipulate the reasoning of the \ac{AI} model by modifying the explanations. This approach becomes particularly effective in the medical context because no technical expertise is required: professionals can redirect and refine the systems' behavior merely by relying on their knowledge and background in their field. This approach enables adaptive learning ensuring the model evolves in alignment with expert reasoning.




Trust calibration is a major challenge in \ac{AI}-assisted decision support, particularly in high-stakes domains like medicine \cite{Kobayashi2024Can}.
More specifically, \ac{AI} systems must be created ensuring that users do not simply accept \ac{AI} recommendations blindly---or, conversely, dismiss them outright \cite{Bach2024Systematic,Mehta2023Artificial,Combi2022Manifesto}. Our proposal tackles this challenge by allowing users to engage with the \ac{AI}'s reasoning process rather than just its final outputs. Instead of simply overriding a classification, experts can refine the reasoning behind it, fostering a symbiotic relationship in which both the user and the \ac{AI} system learn in the process \cite{Desolda2024Humancentered}. Through explanation-driven interventions, users' refinements gradually steer the AI model toward better, safer, and more reliable predictions. This approach positions \ac{AI} as a collaborative partner, learning from domain expertise in a way that strengthens both accuracy and user confidence.


\section{Conclusions and Future Work}\label{sec:conclusions}
This article proposed \emph{explanation-driven interventions} as an \ac{EUD} tool for black-box AI systems, illustrated through their adoption in Rhino-Cyt, a system aiding rhinocytologists in cell counting tasks. Unlike traditional \ac{AI} customization methods that relegate domain experts to passive reviewers, explanation-driven interventions empowers them to directly manipulate \ac{AI}-generated outputs and explanations, without any programming expertise. 
This work contributes to ongoing research in human-centered \ac{AI} and \ac{AI}-assisted decision supports \ac{EUD} systems have the potential to bridge the gap between \ac{AI} automation and expert oversight.
Future work includes evaluating the usability and effectiveness of Rhino-Cyt through user studies, identifying improvement areas, and extending support to multi-user collaboration. Better methods for tracking intervention impact and mitigating user errors are also needed.
Another key direction is assessing how editable explanations affect model performance, accuracy, and user cognitive load. Structured evaluations could clarify whether this approach enhances trust and understanding more than traditional rule-based customization. Finally, explanation-driven interventions hold promise beyond medicine, with potential in domains like law and finance where professionals must interpret and refine AI reasoning. 


\begin{credits}
\subsubsection{\ackname}
This research is partially supported by: 
\begin{enumerate*}[label={(\roman*)}]
    \item the co-funding of the European Union - Next Generation EU: NRRP Initiative, Mission 4, Component 2, Investment 1.3 – Partnerships extended to universities, research centers, companies and research D.D. MUR n. 341 del 15.03.2022 – Next Generation EU (PE0000013 – ``Future Artificial Intelligence Research – FAIR'' - CUP: H97G22000210007); 
    \item  the Italian Ministry of University and Research (MUR) and by the European Union - NextGenerationEU, Mission 4, Component 2, Investment 1.1, under grant PRIN 2022 PNRR ``PROTECT: imPROving ciTizEn inClusiveness Through Conversational AI'' (Grant P2022JJPBY) - CUP: H53D23008150001.
\end{enumerate*}
The research of Andrea Esposito is funded by a Ph.D. fellowship within the framework of the Italian ``D.M. n. 352, April 9, 2022''- under the National Recovery and Resilience Plan, Mission 4, Component 2, Investment 3.3 -- Ph.D. Project ``Human-Centred Artificial Intelligence (HCAI) techniques for supporting end users interacting with AI systems,'' co-supported by ``Eusoft S.r.l.'' (CUP H91I22000410007).
The research of Francesco Greco is funded by a Ph.D. fellowship within the framework of the Italian ``D.M. n. 352, April 9, 2022''- under the National Recovery and Resilience Plan, Mission 4, Component 2, Investment 3.3 -- Ph.D. Project ``Investigating XAI techniques to help user defend from phishing attacks'', co-supported by ``Auriga S.p.A.'' (CUP H91I22000410007).


\subsubsection{\discintname}
The authors have no competing interests to declare that are relevant to the content of this article.
\end{credits}
%
%
%
\bibliographystyle{splncs04}
\bibliography{bibliography}
\end{document}